# First-principles study of electronic and optical properties in wurtzite $Zn_{1-x}Cd_xO$

X. D. Zhang[a]

*Department of Health Physics, Institute of Radiation Medicine, Chinese Academy of Medical Sciences and Peking Union Medical College, Tianjin 300192, China*

M.L. Guo

*Department of Fundamental Subject, Tianjin Institute of Urban Construction, Tianjin 300384, China*

W.X. Li and C.L. Liu

*Department of Applied Physics, School of Science, Tianjin University, Tianjin 300072, China*

## Abstract

A first-principles study has been performed to evaluate the electronic and optical properties of wurtzite $Zn_{1-x}Cd_xO$ up to x=0.25. We have employed the Perdew-Burke-Ernzerhof (PBE) form of generalized gradient approximation within the framework of density functional theory (DFT). Calculations have been carried out in different configurations. With the increasing Cd concentrations, the band gap of $Zn_{1-x}Cd_xO$ is decreased due to the increase of *s* states in conduction band. The results of imaginary part of dielectric function $\varepsilon_2(\omega)$ indicate that the optical transition between O 2*p* states in the highest valence band and Zn 4*s* states in the lowest conduction band has shifted to low energy range as the Cd concentrations increase. Besides, the optical band gap decreases from 3.2 to 2.84 eV with increasing Cd concentrations from 0 to 0.25. Meanwhile, the bowing parameter *b*, which has been obtained by fitting the results of optical band gap, is about 1.21 eV. The optical constants of pure ZnO and $Zn_{0.75}Cd_{0.25}O$, such as optical conductivity, loss function, refractive index and reflectivity, have been discussed.

---

[a] Author to whom correspondence should be addressed;. Tel:+86-22-85682375;
 E-mail: xiaodongzhang@yahoo.cn (X.D. Zhang).

# Ⅰ. INTRODUCTION

ZnO has been used for a variety of applications, such as gas sensors, surface acoustic wave devices, biomedical materials and transparent contacts. Due to direct band gap of 3.37 eV and large binding energy of exciton, ZnO-based semiconductors are recognized as very promising photonic materials in the ultraviolet and visible region. A crucial step in designing modern optoelectronic devices is the realization of band gap engineering to create barrier layers and quantum wells in devices heterostructures. In order to realize such optoelectronic devices, two crucial challenges are *p*-type doping and modulation of band gap. Once *p*-type conducting ZnO is available, band gap engineered electro-optical device will be promising. Thus, modulation of band gap in ZnO is very important. One of interesting features of ZnO is the possibility to tune its band gap by substituting bivalent mentals like Cd or Mg in place of Zn. CdO is a compound semiconductor with direct band gap of 2.3 eV, and Cd substitution can lead to narrowing of band gap depending on the Cd concentrations.[1] So far, ZnCdO alloy has been fabricated by using different doping methods.[2-4] However, band gap narrowing and related optical properties have not been understood completely, and detailed optical transition of ZnCdO alloy is still not clear. Therefore, a first-principles calculation is very interesting to investigate the electronic and optical properties of ZnCdO alloy.

Very recently, the electronic properties of rocksalt CdO, wurtzite ZnO and rocksalt MgO have been investigated by first-principles.[5,6] Furthermore, optical absorption and excitonic properties of wurtzite ZnO have also been detailedly investigated by Laskowski and Christensen.[7] Meanwhile, Some other alloy materials for band gap engineering, such as rocksalt ZnCdO and ZnMgO allloy, have also been investigated by linear muffin tin orbital (LMTO) method, empirical pseudopotential method and full potential linearized augmented plane wave method, respectively.[6,8,9] However, it should be pointed out that the above-mentioned calculations are based on the rocksalt ZnCdO alloy and wurtzite ZnO, and the few investigations were carried out on the wurtzite ZnCdO alloy. The main reason is that wurtzite ZnCdO alloy is difficult to fabricate in earlier investigations,[10] because rocksalt CdO in wurtzite ZnO has the strong solubility limits. However, the problem of solubility limits has been solved by using different doping methods. Many recent investigations have reported that Cd concentrations in wurtzite $Zn_{1-x}Cd_xO$ can be above 50%.[1] Therefore, the theoretical investigations of wurtzite $Zn_{1-x}Cd_xO$ become more and more important. Especially, the first-principles calculations of optical properties of wurtzite $Zn_{1-x}Cd_xO$ are very helpful for understanding its optical transition mechanism.

In present study, we investigate the electronic and optical properties of wurtzite $Zn_{1-x}Cd_xO$ system including ZnO and ZnCdO alloy in different configurations. The paper is organized as follows. The section Ⅱ describes the basic ingredients and details of computational methods we have applied. Section Ⅲ presents and discusses the results of our calculations. First, we calculate the electronic structures, because the optical properties depend on both the interband and intraband transitions, which are determined by energy band. Then, we analyze the optical transition and optical band gap in different configurations. Finally, the optical constants, including the optical reflectivity, conductivity, refractive index and energy loss, have been discussed in wurtzite ZnO and $Zn_{0.75}Cd_{0.25}O$ alloy. Section Ⅳ concludes and summarizes our findings.

## Ⅱ. COMPUTATIONAL METHOD

The calculations are performed with Cambridge Serial Total Energy Package (CASTEP) code, based on density functional theory (DFT) using a plane-wave pseudopotential method.[11] We use the generalized gradient approximation (GGA) in the scheme of Perdew-Burke-Ernzerhof (PBE) to describe the exchange-correlation functional.[12] Previous studies have shown that the norm-conserving pseudopotential, which was developed by hamann el. al,[13] is more suitable than the ultrasoft pseudopotential in optical properties calculations.[14] Thus, norm-conserving pseudopotential is used with $2s^22p^4$ and $3d^{10}4s^2$ as the valence-electron configuration for the oxygen and zinc atoms, respectively, to describe the electron-ion interaction. In this code, the plane wave functions of valence electrons are expanded in a plane wave basis set, and the use of norm-conserving pseudopotential allows a plane wave energy cutoff $E_c$. Only plane waves with kinetic energies smaller than $E_c$ are used in the expansion. Reciprocal-space integration over the Brillouin zone is approximated through a careful sampling at finite number of k-points using a Monkhorst-Pack mesh.[15]

The calculations are performed at constant volume, and the 2×2×2 ZnO supercell has been used, which contains 16 Zn and 16 O atoms, respectively. The substitutional method has been taken into account in this paper, and Cd atoms are used to substitute Zn atoms in ZnO. For x=0.0625, only one Zn atom is substituted by Cd, while x=0.25, four Zn atoms are substituted. In this way, the $Zn_{1-x}Cd_xO$ system is corresponding to Cd concentrations of x=0, 0.0625, 0.125, 0.1875 and 0.25, respectively. We choose the energy cutoff to be 460 eV, and the Brillouin-zone sampling mesh parameters for the k-point set are 9×9×6 and 4×4×2 for pure ZnO and ZnCdO alloy, respectively. The charge densities are converged to $2\times10^{-6}$ eV/atom in the self-consistent calculation. The lattice constant of wurtzite ZnO is 3.24927 Å for $a$ and 5.20544 Å for $c$, respectively.[16] The density of wurtzite $Zn_{1-x}Cd_xO$ system is 5.67848, 5.88358, 6.08869, 6.29379 and 6.49890 g/cm$^3$, respectively, which is corresponding to different Cd concentrations from x=0 to x=0.25. Recent investigations from cation doped ZnO indicate that the distance of doping atoms can influence calculated results due to interaction of doping atoms.[17] So, for weakening the interaction with Cd-Cd, the separation distance is as far as possible. And then, the $Zn_{1-x}Cd_xO$ system is optimized with lattice constants and the positions of substitutional atoms. In the optimization process, the energy change, maximum force, maximum stress and maximum displacement tolerances are set as $2\times10^{-5}$ eV/atom, 0.05 eV/Å, 0.1 Ga and 0.002 Å, respectively. Besides, for obtaining exact optical absorption spectra in low energy range, the scissors operation has been carried out in optical absorption of ZnO and ZnCdO alloy.

## Ⅲ. RESULTS AND DISCUSSIONS

**A. Electronic properties**

Fig. 1 shows the band structure of pure ZnO. It is observed that the direct band gap is about 0.5 eV at highly symmetric Γ point, which is close to earlier calculated result.[18] It is well known that the underestimated band gap can be due to the choice of exchange-correlation energy. For pure ZnO, the band gap calculated with LDA or GGA is about 0.5-1.0 eV, and the band gap calculated by GGA is always less than that of LDA. The valence band mainly consists of the 2*p*, 2*s* states of O and 3*d* states of Zn. In the uppermost valence band, O 2*p* states are predominantly found between -4 and 0 eV. While the O 2*s* states appear in the range from -18.5 to -16.5eV, which are very close to the previous calculated results.[18] It is clear that Zn 3*d* states give rise to some bands in the energy range of 4-6 eV below the valence band maximum (VBM), which shows a splitting and a wave-vector dispersion outside Γ. In addition, the lowest conduction band is dominated by Zn 4*s* states, and experimental results also indicate that it is primarily derived from O 2*p* and Zn 4*s* states.[19]

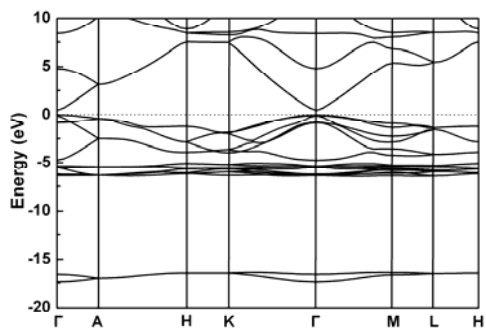

**Fig.1 Band structure of pure ZnO**

It should be noticed that the huge peak in valence band (at Γ) caused by these basically Zn 3*d* states is observed. Moreover, the Zn 3*d* states act more subtle on band structure via the repulsion of *p* and *d* states, which is caused by hybridization of respective states. Thus, underestimation of band gap is further enhanced in ZnO due to the hybridization between the Zn 3*d* and O 2*p* levels. The earlier calculations indicate that it is a contribution of 20-30% of Zn 3*d* states to the levels in the upper valence band,[20] which is more than 9% of the experimental result.[21]

Fig.2 gives the density of states (DOS) of $Zn_{1-x}Cd_xO$ system. Compared with the pure ZnO in Fig.2 (a), the Cd 4*d* states in ZnCdO alloy have been observed in the energy range of 7-8 eV below VBM [Fig.2(b)-(e)]. Meanwhile, the Cd 4*d* states enhance gradually with the increasing Cd concentrations, while the Zn 3*d* states between -7 and -5 eV decrease slightly. With the increasing Cd concentration, O 2*s* states have slightly shifted to low energy range. Besides, it can be observed that both O 2*p* and Zn 3*d* states become more delocalized in the top of valence band with a narrower bandwidth as the Cd concentrations increase. As a result, the excitons will be localized effectively in ZnCdO alloy, which can influence its band edge emission.

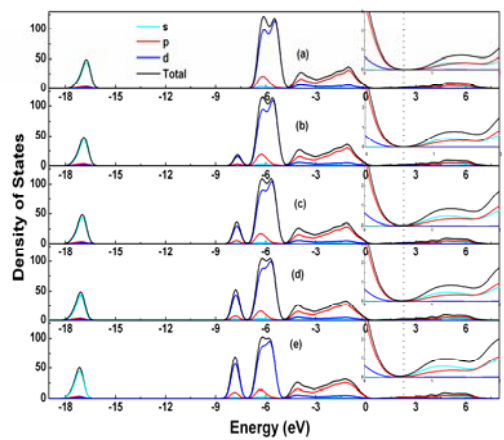

**Fig.2 The partial DOS of $Zn_{1-x}Cd_xO$ when (a) =0; (b) x=0.0625; (c) x=0.125; (d) x= 0.1875; and (e) x=0.25. Each inset is the magnified spectrum in the energy range of 0-2 eV.**

It is clearly seen from Fig.2 (b) to 2(e) that when x changes from 0.0625 to 0.25, the DOS also changes. In general, the band gap of ZnCdO alloy is less than that of pure ZnO. It confirms that the Cd incorporation into ZnO can induce the obvious band gap narrowing, which is consistent with the previous experimental results.[1,2] The exact band gap of

$Zn_{1-x}Cd_xO$ are 0.30, 0.20, 0.16, and 0.13 eV, which are corresponding to the different x values from 0.0625 to 0.25. It is necessary to point out that the VBM is always in Fermi level and it has no obvious shift, which can be clearly seen in the inset which is the magnified spectrum in the energy range of 0-2 eV. The similar investigation has been carried out in ZnMgO alloy.[22] In the bottom of conduction band, when x increases from 0.0625 to 0.25, the conduction band has shifted to low energy range. This shift can lead to the obvious variation of electronic properties. It is well known that the bottom of conduction band consists of Zn 4$s$ and O 2$p$ states, and the Zn 4$s$ states are dominant. Cd incorporation can enhance s states. As it is incorporated more and more, the s states at the bottom of conduction band become stronger and stronger, which leads to conduction band shift and band gap narrowing. It is noting that the variations of band gap are not obvious in high concentration ($\geq$12.5%), and it may indicate that the modulation of band gap by Cd substitution is limited. This may be related to the band gap of 2.3 eV in CdO. Experimentally, the band gap of $Zn_{1-x}Cd_xO$ alloy can be modulated in the range of 2.65-3.3 eV with the Cd concentrations from x=0 to x=0.53.[23]

**B. Optical properties**

For investigating the optical band gap and optical transition of the $Zn_{1-x}Cd_xO$ system, it is necessary to investigate the imaginary part of the dielectric function $\varepsilon_2(\omega)$, because $\varepsilon_2(\omega)$ is very important for optical properties of any materials. It is well known that the interaction of a photon with the electrons in the system can be described in terms of time-dependent perturbations of the ground-state electronic states. Optical transitions between occupied and unoccupied states are caused by the electric field of the photon. The spectra from the excited states can be described as a joint density of states between the valence and conduction band. The momentum matrix elements, which are used to calculate the $\varepsilon_2(\omega)$, are calculated between occupied and unoccupied states which are given by the eigen vectors obtained as solution of the corresponding Schrödinger equation. Evaluating these matrix elements, one uses the corresponding eigen functions of each of the occupied and unoccupied states. The real part of dielectric function $\varepsilon_1(\omega)$ can be evaluated from the imaginary part $\varepsilon_2(\omega)$ by the famous Kramer-Kronig relationship. Absorption coefficient $\alpha(\omega)$ can be obtained from the $\varepsilon_1(\omega)$ and $\varepsilon_2(\omega)$,

$$\alpha(\omega) = \sqrt{2}\omega[\sqrt{\varepsilon_1^2(\omega) + j\varepsilon_2^2(\omega)} - \varepsilon_1(\omega)]^{1/2}.$$

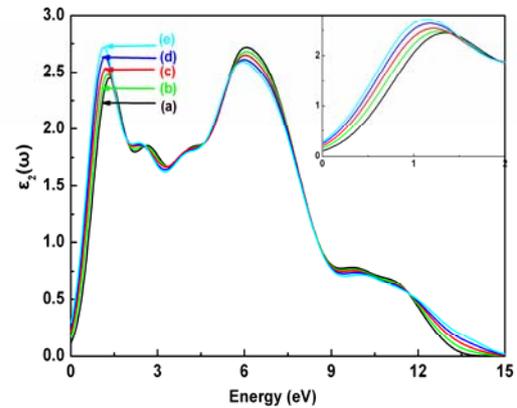

**Fig.3 The imaginary part of dielectric function $\varepsilon_2(\omega)$ of $Zn_{1-x}Cd_xO$ when (a) =0; (b) x=0.0625; (c) x=0.125; (d) x= 0.1875; and (e) x=0.25.**

Fig.3 gives the imaginary part of dielectric function $\varepsilon_2(\omega)$ of pure ZnO and ZnCdO alloy. To the pure ZnO [Fig.3 (a)], there are three main peaks in $\varepsilon_2(\omega)$, which are 1.35, 6.10 and 10.6 eV, respectively. The peak at 1.35 eV should mainly be caused by optical transitions between O 2$p$ states in the highest valence band and Zn 4$s$ states in the lowest conduction band, which is very close to the other first-principles evaluations (1.4 eV),[18] and it is not far from the direct band gap of 0.5 eV. The peak at 6.10 eV can be due to optical transition between the Zn 3$d$ and O 2$p$ states, and the weak peak at 10.6 eV is mainly derived from the optical transition between the Zn 3$d$ and O 2$s$ states. The two peaks are very close to the other estimations.[5] The $\varepsilon_2(\omega)$ of ZnCdO alloy can be found in the Fig.3 (b)-(e). The optical transition between O 2$p$ and Zn 4$s$ states has been affected by Cd incorporation. The optical transitions between the Zn 4$s$ and O 2$p$ states are 1.28, 1.21, 1.16, and 1.14 eV, respectively, which are corresponding to the different Cd concentrations from x=0.0625 to 0.25. The shift of the optical transition indicates that the direct band gap is decreasing, which is in good agreement with the result of DOS. Meanwhile, the optical transition has enhanced gradually with the increasing Cd concentrations owing to increasing $s$ states in valence band, which may induce the enhancement of the band edge emission of ZnCdO alloy. However, the recent photoluminescence results indicate the band edge emission of Zn$_{1-x}$Cd$_x$O has no obvious variation.[1,24] This may be ascribed to localized excitons in ZnCdO alloy, and related investigations are still necessary. Besides, with the increasing Cd concentrations, it can be clearly seen that the peak at 10.6 eV decreases slightly, which can be ascribed to the decreasing Zn 3$d$ states. Meanwhile, the optical transition between the Zn 3$d$ and O 2$p$ states (~6.1 eV) has shifted to low energy range with slight decrease, which is due to the more and more localized and decreasing Zn 3$d$ states.

Fig.4 gives the optical absorption spectra of Zn$_{1-x}$Cd$_x$O system under scissors operation in the range of 0-5 eV. Due to the underestimation of band gap, it is difficult to obtain the exact optical band gap. In our calculations, we have used the energy scissors approximation with 2.7 eV to fit the absorption edge to experimental value. This method is effective for a variety of systems.[25] It is clearly observed that the Cd incorporation induces the redshift of optical absorption edge. It is well known that relation between optical band gap and the absorption coefficient is given by[26]

$$\alpha h\nu = c(h\nu - E_g)^{1/2},$$

where $h$ is Planck's constant, $c$ is a constant for a direct transition, $\nu$ is frequency of radiation and $\alpha$ is the optical absorption coefficient. The optical band gap $E_g$ can be obtained from the intercept of $(\alpha h\nu)^2$ versus photon energy ($h\nu$). By using the extrapolation, optical band gap of ZnO and ZnCdO alloy can be obtained. The optical band gap depending on Cd concentration has been shown in the inset. It can be observed that the optical band gap decreases from 3.20 to 2.84 eV with the increasing Cd concentrations, which is in good agreement with the previous experimental investigations.[1-4] Especially, optical band gap $E_g$ of ternary semiconductor Zn1-xCdxO is determined by following equation:[27]

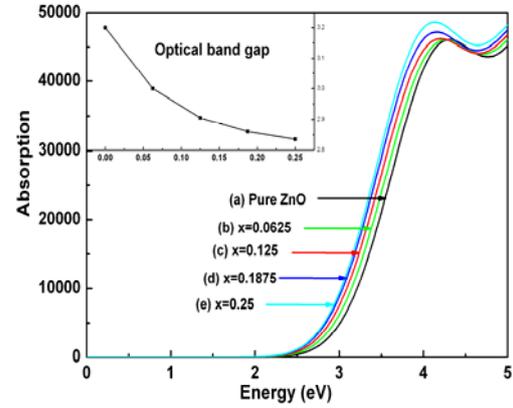

**Fig.4 Optical absorption spectra of Zn$_{1-x}$Cd$_x$O when (a) =0; (b) x=0.0625; (c) x=0.125; (d) x= 0.1875; and (e) x=0.25. Inset is Cd concentration-dependent optical band gap.**

$$E_g = (1-x)E_{ZnO} + xE_{CdO} - bx(1-x),$$

where b is the bowing parameter, and $E_{CdO}$ and $E_{ZnO}$ are the band gap of compounds CdO and ZnO, respectively. The bowing parameter b depends on the difference in electronegativities of the end binaries ZnO and CdO. Suitable bowing parameter is very important for the accuracy of band gap engineering. By fitting the calculated optical band gap, the bowing parameter *b* of 1.21 eV has been obtained, which is less than the previous reported value of 8.14eV and 5.95 eV.[3,10] However, it is not far from the recent result of 0.95 eV, which is calculated by optical reflectivity and absorption data.[28] Wang et al [28] have pointed out that the discrepancy may be ascribed to the ignoring the excitonic contributions of optical absorption edge in previous studies. In addition, the optical band tail, which is induced by the DOS of localization, may also influence the bowing parameter b.[29]

Fig 5 gives the optical constants of pure ZnO and $Zn_{0.75}Cd_{0.25}O$ alloy in the range of 0-15 eV. The optical constants, such as optical conductivity, loss function, refractive index and reflectivity, are very important for the optical materials and related applications. Overall, the optical constants of ZnCdO alloy have changed slightly. For example, the effect on conductivity [Fig.5 (a)] by Cd incorporation is not obvious in the low energy range, while it decreases sharply above 12 eV, which is corresponding to variation of $\varepsilon_2(\omega)$. Similar result has also been obtained in loss function spectra [Fig.5 (b)], which can describe the energy loss of fast electron traversing in $Zn_{0.75}Cd_{0.25}O$. The peaks of 9.0 and 12.6 eV in loss function are plasma resonance peaks, and they are also the points of transition from the metallic property to the dielectric properties for $Zn_{0.75}Cd_{0.25}O$. The Cd incorporation induces the blueshift of 12.6 eV peak with sight decrease in loss function spectra. The knowledge of refractive index of $Zn_{0.75}Cd_{0.25}O$ alloy is necessary for accurate modeling and design of devices. The refractive index [Fig.5 (c)] and reflectivity [Fig.5 (d)] of $Zn_{0.75}Cd_{0.25}O$ are increased in low energy range, which indicate that the band gap decreases.[28] In particular, the maximum of refractive index of pure ZnO in the low energy range is 2.3, which is very close to the experimental data of 2.1 by optical spectra.[30,31] However, it is not as sharp as experimental results. The peak is closely related to the optical transition near the band gap. The refractive index of $Zn_{0.75}Cd_{-0.25}O$ would increase in low energy range, which is contrary to that of ZnMgO alloy[32]. It is not difficult to understand, because the Mg incorporation into ZnO can induce the decrease of Zn 4$s$ states in the bottom of conduction band, while Cd incorporation into ZnO can lead to enhancement of $s$ states in conduction band. As a result, the optical transition between the uppermost valence band and the lowest conduction band can increase in $Zn_{0.75}Cd_{-0.25}O$ alloy.

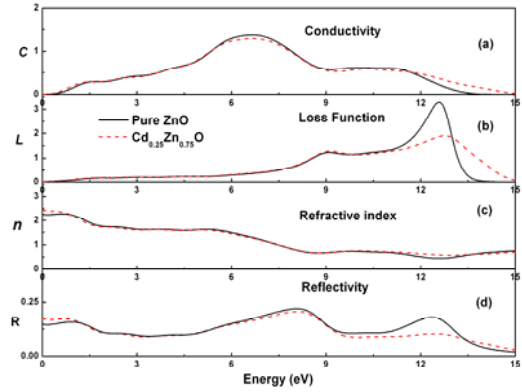

**Fig.5 Optical conductivity (a), loss function (b), refractive index (c) and optical reflectivity (d) of pure ZnO and $Zn_{0.75}Cd_{0.25}O$.**

## IV. CONCLUSION

In summary, a first-principles study has been performed to evaluate the electronic and optical properties of $Zn_{1-x}Cd_xO$ system in different configurations. The increasing Cd concentration can induce the band gap narrowing, and the band gap of $Zn_{1-x}Cd_xO$ system has decreased from 0.5 to 0.13 eV. Subsequently, optical transition between uppermost valence band and the lowest conduction band has shifted to low energy range with the increasing Cd concentration. Besides, the optical band gap has decreased from 3.2 to 2.84 eV with increasing Cd concentrations from 0 to 0.25. Meanwhile, the bowing parameter $b$ is about 1.21 eV. The Cd incorporation induces the slight variations of optical constants. The refractive index of $Zn_{0.75}Cd_{0.25}O$ alloy is higher than that of pure ZnO in low energy range.

**ACKNOWLEDGMENTS**

Authors would like to thank Prof. Hui-Tian Wang (Nanjing University), Prof. Mike Payne (University of Cambridge), and Prof. Sheng-Bai Zhang (National Renewable Energy Laboratory) for some helpful suggestions and discussions. This work is supported by the Natural Science Foundation of Tianjin (Grant No. 06YFJMJC01100) and National Natural Science Foundation of China (Grant No.10675089).